\documentclass[11pt]{amsart}
\usepackage{amsmath,amssymb} 
\usepackage{graphicx}
\usepackage[utf8]{inputenc} 
\usepackage[font=small,labelfont=bf]{caption}
\usepackage{bbm}

\begin{document}

\newtheorem{thm}{Theorem}[section]
\newtheorem{lem}[thm]{Lemma}
\newtheorem{prop}[thm]{Proposition}
\newtheorem{coro}[thm]{Corollary}
\newtheorem{defn}[thm]{Definition}
\newtheorem*{remark}{Remark}

\numberwithin{equation}{section}

\newcommand{\Z}{{\mathbb Z}} %cph changed from \mathbf
\newcommand{\Q}{{\mathbb Q}}
\newcommand{\PP}{{\mathbb P}}
\newcommand{\R}{{\mathbb R}}
\newcommand{\C}{{\mathbb C}}
\newcommand{\N}{{\mathbb N}}
\newcommand{\FF}{{\mathbb F}}
\newcommand{\T}{{\mathbb T}}
\newcommand{\fq}{\mathbb{F}_q}

\newcommand{\fixmehidden}[1]{}

\def\scrA{{\mathcal A}}
\def\cB{{\mathcal B}}
\def\cC{{\mathcal C}}
\def\Eps{{\mathcal E}}
\def\cI{{\mathcal I}}
\def\scrD{{\mathcal D}}
\def\cF{{\mathcal F}}
\def\cL{{\mathcal L}}
\def\cM{{\mathcal M}}
\def\cN{{\mathcal N}}
\def\cP{{\mathcal P}}
\def\scrR{{\mathcal R}}
\def\scrS{{\mathcal S}}
\def\j{\mathfrak{j}}

\newcommand{\rmk}[1]{\footnote{{\bf Comment:} #1}}

\renewcommand{\mod}{\;\operatorname{mod}}
\newcommand{\ord}{\operatorname{ord}}
\newcommand{\TT}{\mathbb{T}}
\renewcommand{\i}{{\mathrm{i}}}
\renewcommand{\d}{{\mathrm{d}}}
\renewcommand{\^}{\widehat}
\newcommand{\HH}{\mathbb H}
\newcommand{\Vol}{\operatorname{vol}}
\newcommand{\area}{\operatorname{area}}
\newcommand{\tr}{\operatorname{tr}}
\newcommand{\norm}{\mathcal N} % norm =(\frac{ n+\sqrt{n^2-4}} 2)^2
\newcommand{\intinf}{\int_{-\infty}^\infty}
\newcommand{\ave}[1]{\left\langle#1\right\rangle} %  average
\newcommand{\E}{\mathbb E}
\newcommand{\Var}{\operatorname{Var}}
\newcommand{\Cov}{\operatorname{Cov}}
\newcommand{\Prob}{\operatorname{Prob}}
\newcommand{\sym}{\operatorname{Sym}}
\newcommand{\disc}{\operatorname{disc}}
\newcommand{\CA}{{\mathcal C}_A}
\newcommand{\cond}{\operatorname{cond}} % conductor
\newcommand{\lcm}{\operatorname{lcm}}
\newcommand{\Kl}{\operatorname{Kl}} %Kloosterman sum
\newcommand{\leg}[2]{\left( \frac{#1}{#2} \right)}  % Legendre symbol
\newcommand{\id}{\operatorname{id}}
\newcommand{\beq}{\begin{equation}}
\newcommand{\eeq}{\end{equation}}
\newcommand{\bsp}{\begin{split}}
\newcommand{\esp}{\end{split}}
\newcommand{\bra}{\left\langle}
\newcommand{\ket}{\right\rangle}
\newcommand{\diam}{\operatorname{diam}}
\newcommand{\supp}{\operatorname{supp}}
\newcommand{\dist}{\operatorname{dist}}
\newcommand{\sgn}{\operatorname{sgn}}
\newcommand{\inte}{\operatorname{int}}
\newcommand{\re}{\operatorname{Re}}

\newcommand{\sumstar}{\sideset \and^{*} \to \sum}

\newcommand{\LL}{\mathcal L} %L-function of u
\newcommand{\sumf}{\sum^\flat}
\newcommand{\Hgev}{\mathcal H_{2g+2,q}}
\newcommand{\USp}{\operatorname{USp}}
\newcommand{\conv}{*}
\newcommand{\CF}{c_0} % Fejer constant
\newcommand{\kerp}{\mathcal K}

\newcommand{\gp}{\operatorname{gp}}
\newcommand{\Area}{\operatorname{Area}}

%\title[No random waves in \u{S}eba's billiard]{No random waves in
%\u{S}eba's billiard} 
\title[Multifractal eigenfunctions for quantum star graphs] 
{Multifractal eigenfunctions \\ for quantum star graphs} 
  
\author{Jonathan P. Keating and Henrik Uebersch\"ar}
\address{Sorbonne Universit\'e, Universit\'e de Paris, CNRS, IMJ-PRG, F-75006 Paris, France.}
\email{henrik.ueberschar@cnrs.fr}
\address{Mathematical Institute, University of Oxford, Andrew Wiles Building, Radcliffe Observatory Quarter, Woodstock Road, 
Oxford OX2 6GG, UK}
\email{jon.keating@maths.ox.ac.uk}
\date{\today}

\date{\today}
\maketitle
 
\begin{abstract}
We prove that the eigenfunctions of quantum star graphs exhibit multifractal self-similar structure in certain specified circumstances. In the semiclassical regime, when the spectral parameter and the number of vertices tend to infinity, we derive an asymptotic condition for the Mellin transform of a specified function arising from the set of bond lengths which yields an asymptotic for the Renyi entropy associated with an eigenfunction. We apply this result to show that one may construct simple quantum star graphs which satisfy a multifractal scaling law. In the low frequency regime we  prove multifractality by computing the Renyi entropy in terms of a zeta function associated with the set of bond lengths. In certain arithmetic cases the fractal exponent $D_q$ satisfies a symmetry relation around $q=1/4$ which arises from the functional equation of the zeta function. Our results are, in some sense, analogous to the multifractal scaling law that the authors recently proved for arithmetic \u{S}eba billiards.  However, unlike in that case, we do not require arithmetic conditions to be satisfied, and nor do we rely on delicate arithmetic estimates. 
\end{abstract}

\section{Introduction}

For the most part, the mathematical literature on quantum chaos has focused on topics such as quantum (unique) ergodicity and scarring of eigenfunctions \cite{A08,CdV85,Z87,L06,So10,FNdB03,Ha10,Ho10a,Ho10b}; however, it is believed that some important classes of quantum systems possess eigenfunctions whose morphology is far more complex than being simply ergodic or localized. For example, in systems whose spectral statistics is intermediate between random-matrix and Poisson there is a longstanding expectation that the eigenfunctions should exhibit multifractal self-similar structure.

One such class of intermediate systems are those termed pseudointegrable \cite{RB81}. Their classical dynamics is close to integrable, yet their quantum dynamics displays features similar to, but distinct from, that usually observed for chaotic systems \cite{BoGeSch01}. Examples are rational polygonal billiards, in which the billiard flow can be studied via the geodesic flow on flat surfaces with conical singularities. Often included in this class are toy models of such systems, such as \u{S}eba billiards \cite{Se90, BoLeSch00}, quantum star graphs \cite{BeBoK01,BK99,BKW03,BKW04} and parabolic toral automorphisms \cite{MR00}.

Multifractality is ubiquitous in many areas of science (biology, financial mathematics, quantum physics, etc). It describes self-similarity that is so complex it cannot be captured by a single fractal exponent.  Instead, it can only be characterised by a spectrum of exponents. While multifractality is an important concept in the theoretical and experimental study of intermediate quantum systems, and evidence for its existence has been found in a wide range of examples \cite{Al86,KM97,AtBo12,BHK19,Gi20}, there are relatively few rigorous mathematical results relating to it in that context.

We recently proved \cite{KU21} the existence of multifractal eigenfunctions for \u{S}eba billiards, namely rectangular billiards with a Dirac delta potential. To the best of our knowledge, this is the first rigorous proof of quantum multifractality, at least in the context of quantum chaos. In particular, we obtained a multifractal scaling law for the Renyi entropy in the semiclassical limit for the case of arithmetic tori, as well as for the ground state in the case of general rectangular tori. Moreover, we discovered that a symmetry relation predicted in the physics literature for the fractal exponent manifests itself, in the case of the ground state, as the functional equation satisfied by Epstein's zeta function. 

In the present article we deal with another toy model of pseudointegrable dynamics: quantum star graphs. These systems have been extensively studied in the mathematical literature on quantum chaos \cite{BeBoK01,BK99,BKW03,BKW04}. Star graphs are graphs with a central vertex that is connected to $v$ outlying vertices. One can define a self-adjoint operator by introducing the $1d$ Laplacian on each bond and imposing suitable boundary conditions at the central vertex \cite{KS99} (cf. section \ref{graphs}). Taking the semiclassical limit, as the eigenvalue tends to infinity, and at the same time $v\to+\infty$, one can show that the spectral equation and expressions for the eigenfunctions are of a form which closely resembles the analogous expressions for \u{S}eba billiards. Indeed the statistical properties of the eigenvalues and eigenfunctions in the two systems are closely related \cite{BeBoK01,KU17}. 

While our proof of multifractality for \u{S}eba billiards in the semiclassical regime was limited to arithmetic billiards and relied heavily on arithmetic methods, we are able here to obtain analogous results for a generic class of quantum star graphs. We construct examples which give rise to a multifractal scaling law for the Renyi entropy. In the case of the ground state regime we derive a general multifractal scaling law and compute the fractal exponent in terms of a zeta function defined in terms of the set of bond lengths. In certain star graphs, the symmetry relation satisfied by the fractal exponent manifests itself via the functional equation for the associated zeta function.

{\bf Acknowledgements.} JPK is pleased to acknowledge support from ERC Advanced Grant 740900 (LogCorRM).  H.U. was partially supported by the grant ANR-17-CE40-0011-01 of the French National
Research Agency ANR (project SpInQS) and by a ``d\'el\'egation CNRS''.   

\section{Background and main results}

We will discuss the spectral properties of quantum graphs in more detail in section \ref{sec-graphs}. Here we simply introduce what is necessary to formulate our results.
Let us consider a star graph with bond lengths $L_1,\cdots,L_v$. 

The spectrum of the self-adjoint operator defined on the star graph forms a discrete set
$$0\leq k_0^2<k_1^2\leq k_2^2\leq \cdots\leq k_j^2\leq \cdots\to+\infty$$
where the parameters $\pm k_j$ are solutions of a spectral equation. 
are solutions of the spectral equation
\beq
Z(k,L):=\sum_{j=1}^v \tan(kL_j)=\frac{1}{\alpha k}
\eeq
where $\alpha$ denotes a coupling parameter. One may consider a renormalized coupling regime, where $\alpha=\alpha(k)$ may depend on the spectral parameter $k$.

In particular, the $k_j$ interlace with the poles of $Z(k,L)$
which are given by $$s_{m,j}:=(m+\frac{1}{2})\pi/L_j, \quad m\in\Z;\; j=1,\cdots,v.$$
Moreover, we have the ordering
$$s_{m,1}\geq s_{m,2}\geq\cdots\geq s_{m,v}.$$

Let us assume $L_j=L + \epsilon(v)\ell_j$, where $\ell_1\leq\ell_2\leq\cdots\leq\ell_v$. Then we have
$$s_{m,j}=(m+\frac{1}{2})\frac{\pi}{L}+O(\frac{m\epsilon\ell_j}{L^2})
=(m+\frac{1}{2})\frac{\pi}{L}+O(\frac{m\epsilon\ell_v}{L^2}).$$ 
In particular, one has $$|s_{m,j}-s_{m,k}|\asymp \frac{m\epsilon}{L^2}|\ell_j-\ell_k|.$$

We will consider the semiclassical limit $k\to+\infty$ and at the same time we will take $v\to+\infty$ in such a way that $k_n \epsilon(v)\ell_v\to0$. This ensures $s_{m,j}=(m+1/2)\pi/L+o(1)$, and because the $k_j$ interlace with the poles of the spectral function it follows that they form clusters around the values $(m+1/2)\pi/L$ which are well separated. So with each $m\in\Z$ we may associate a cluster $\cC_m$.

Moreover, there are precise formulae for the eigenfunctions, and for the maximum amplitude squared on each bond. 
We, thus, obtain a probability measure on the finite set $\Omega_v=\{1,\cdots,v\}$ given by (cf. subsection \ref{def-Renyi})
\beq
\mu_k(i)=\frac{\sec(k L_i)^2}{\sum_{j=1}^v \sec(k L_j)^2}.
\eeq

The Renyi entropy associated with the probability measure $\mu$ on $\Omega_v$ is then given by
\beq
H_q(\mu_k)=\frac{1}{1-q}\log\left(\sum_{j=1}^v \mu_k(j)^{q}\right). 
\eeq

Let us define the Mellin transform of a compactly supported piecewise continuous function $f$ as
$$(\cM f)(s)=\int_0^\infty t^{-s}f(t)dt.$$
Let $\j\in C^1(\R_+)$ be an invertible function which satisfies $\j(\ell_j)=j$, $\forall j=1,\dots,v$.
We introduce the length function $$F(v,\j)(t)=(\mathbbm{1}_{[\ell_{i+2},\ell_v]}\j')(t+\ell_{i+1})$$

We have the following result which provides a condition in terms of the Mellin transform of the length function to ensure that an asympotic of the Renyi entropy may be obtained.
\begin{thm}
If $$\lim_{v\to+\infty}\frac{\log (\cM F(v+1,\j))(s)}{\log (\cM F(v,\j))(s)}=1,$$ then the asymptotics of the Renyi entropy may be computed as 
\begin{equation}
H_q(\mu_k)\sim\frac{\log (\cM F(v,\j))(2q)-q\log (\cM F(v,\j))(2)}{1-q}.
\end{equation}
\end{thm}

We may apply the theorem to deduce the following corollary which shows that one may construct simple quantum star graphs with a multifractal scaling law.
\begin{coro}
Let $\j(t)=t^n$, $n\in\N$. Denote by $k_m$ the first eigenvalue inside the cluster $\cC_m$ for $m\in\N$ and by $\sigma_m$ the distance to the nearest pole. Assume that
$$\liminf_{m\to\infty}\sigma_m>0.$$

For $q>n/2$ and $n\geq 2$, we may compute the fractal exponent as
\beq
D_q=\lim_{k\to+\infty}\frac{H_q(\mu_k)}{\log v}=
\begin{cases}
1-\frac{2q-n}{n(q-1)}, \quad\text{if}\; n\geq3,\\
\\
0, \quad\text{if}\; n=2.
\end{cases}
\end{equation}
\end{coro} 

Next we state our results in the ground state regime.  These are in some sense analogous to the results we obtained for \u{S}eba's billiard, however, a wide variety of zeta functions may appear. Note that in the regime of small eigenvalues the width of the clusters does not blow up with $k$, and therefore we do not have to take $v\to+\infty$ at the same time, as we vary the spectral parameter. Rather, let us suppose that $\epsilon(v)\ell_v$ is small but fixed. Then we are in the clustering regime and we may compute the Renyi entropy for any fixed $v$.
\begin{thm}
Let us denote by $k_0^2$ the lowest eigenvalue associated with the cluster $\cC_0$ and we write $k_0=\pi/(2L)+\sigma_0$.
For $v$ fixed, we have
\beq
D_{q}(v)=\lim_{\sigma_0\to 0}H_q(\mu_{k_0})=\frac{\log\zeta_{G,v}(2q)-q\log\zeta_{G,v}(2)}{1-q}
\eeq
where 
\beq
\zeta_{G,v}(s)=\sum_{j=1}^v\left(\frac{\ell_j}{L+\epsilon(v)\ell_j}\right)^{-s}
\eeq

If $\re q>1/2$, and $\epsilon(v)\ell_v\to 0$, as $v\to\infty$, then
\beq
D_q=\lim_{v\to\infty}D_{q}(v)=\frac{\log\zeta_{G}(2q)-q\log\zeta_{G}(2)}{1-q},
\eeq 
where
\beq
\zeta_{G,v}(s)=\sum_{j=1}^\infty\ell_j^{-s}.
\eeq
\end{thm} 

Depending on the choice of the set $S=\{\ell_j\}_{j=1}^\infty$, which corresponds to a star graph $G(S)$, different types of zeta functions may arise which may satisfy a functional equation. For example $S=\N$ gives rise to Riemann's zeta function 
$$\zeta_{G(\N)}(s)=\zeta(s)=\sum_{n=1}^{\infty}n^{-s}$$
whereas $S=\cN=\{|\xi|^2 \mid \xi\in\cL\}$, where $\cL=\Z(1,0)\oplus\Z(0,\sqrt{D})$ denotes a rectangular lattice, gives rise to Epstein's zeta function associated with the quadratic form $Q(x,y)=x^2+Dy^2$,
$$\zeta_{G(\cN)}(s)=\zeta_Q(s)=\sum_{(x,y)\in\Z^2}Q(x,y)^{-s}.$$

In these cases, the functional equation for the respective zeta function 
$$\zeta_G(1-s)=\varphi_G(s)\zeta_G(s)$$ 
yields a symmetry relation around the critical point $q=1/4$:
\beq 
D_{1/2-q}=D_q\times\frac{1-q}{1/2+q}+\frac{\log\varphi(2q)+(2q-1/2)\log\zeta(2)}{1/2+q}
\eeq

\section{Spectral theory of quantum star graphs}\label{sec-graphs}

In this section we briefly recall the definition and spectral properties of quantum star graphs which are necessary to introduce quantities such as the Renyi entropy in this context. For a more detailed discussion we refer the reader to the introduction of the paper \cite{KMW03}.

\subsection{Schr\"odinger operator on star graphs}\label{graphs}
A star graph $G_v$ consists of a central vertex and $v$ outlying vertices. There are, thus, $v$ bonds and we denote the associated bond lengths by $L_1,\cdots,L_v$. We denote the vector of bond lengths by $L=(L_1,\cdots,L_v)$. Moreover we assume these to be ordered according to increasing length:
$$L_1\leq L_{2}\leq\cdots\leq L_v.$$

Let us introduce a space of $C^2$ functions $\psi$ on $G_v$, where we denote by $\psi_j$ the component of the function on the $j$th bond. We identify each bond with an interval $[0,L_j]$ and introduce a coordinate $x$ on the interval. Then $\psi_j:[0,L_j]\mapsto\R$. For each bond the point $x=0$ corresponds to the central vertex, whereas $x=L_j$ corresponds to the outlying vertex respectively.

The Schr\"odinger operator on $G_v$ is given by the Euclidean Laplacian $-d^2/dx^2$ and the components of the function $\psi$ must satisfy the following matching conditions:
\beq
\psi_i(0)=\psi_j(0)=C, \quad \text{for all}\; i,j=1,\cdots,v,
\eeq
where $C$ is a real constant,
and 
\beq
\sum_{j=1}^v\psi_j'(0)=\frac{C}{\alpha},
\eeq
as well as 
\beq
\psi_j'(0)=0, \quad \text{for all}\; j=1,\cdots,v,
\eeq
where $\alpha\in\R\setminus\{0\}$ denotes a coupling parameter \cite{KS99}.

%\subsection{Spectrum}
%The above defined operator is self-adjoint and its eigenvalues form a discrete set
%$$0\leq k_0^2<k_1^2\leq k_2^2\leq \cdots\leq k_n^2\leq \cdots\to+\infty$$
%which accumulates at infinity and the parameters $\pm k_j$
%are solutions of the spectral equation
%\beq
%Z(k,L):=\sum_{j=1}^v \tan(kL_j)=\frac{1}{\alpha k}.
%\eeq
%In particular, they interlace with the poles of $Z(k,L)$
%which are given by $$s_{m,j}:=(m+\frac{1}{2})\pi/L_j, \quad m\in\Z;\; j=1,\cdots,v.$$
%In particular, we have the ordering
%$$s_{m,1}\geq s_{m,2}\geq\cdots\geq s_{m,v}.$$

%Let us assume $L_j=L + \epsilon(v)\ell_j$, where $\ell_1\leq\ell_2\leq\cdots\leq\ell_v$. Then we have
%$$s_{m,j}=(m+\frac{1}{2})\frac{\pi}{L}+O(\frac{m\epsilon\ell_j}{L^2})
%=(m+\frac{1}{2})\frac{\pi}{L}+O(\frac{m\epsilon\ell_v}{L^2}).$$

%We will consider the semiclassical limit $k_n\to+\infty$ and at the same time we will take $v\to+\infty$ in such a way that $k_n \epsilon(v)\ell_v\to0$. This ensures $s_{m,j}=(m+1/2)\pi/L+o(1)$, and because the $k_n$ interlace with the poles of the spectral function it follows that they form clusters around the values $(m+1/2)\pi/L$ which are well separated.

\subsection{Renyi entropy for star graphs}\label{def-Renyi}

As explained in \cite{KMW03}, it is natural to study the properties of the eigenfunctions of a quantum star graph in terms of the maximum amplitude squared on each bond length. This can be calculated explicitly for an $L^2$ normalized eigenfunction with spectral parameter $k$ on the $i$th bond as
\beq
A_i(k,L;v)=\frac{2\sec(k L_i)^2}{\sum_{j=1}^v L_j\sec(k L_j)^2}.
\eeq

This gives rise to a probability measure on the finite set $\Omega_v=\{1,\cdots,v\}$, given by
\beq
\mu_k(i)=\frac{\sec(k L_i)^2}{\sum_{j=1}^v \sec(k L_j)^2}.
\eeq

The Renyi entropy associated with $\mu$ on $\Omega_v$ is then given by
\beq
H_q(\mu_k)=\frac{1}{1-q}\log\left(\sum_{i=1}^v \mu(i)^{q}\right). 
\eeq

We may rewrite this as 
\beq
H_q(\mu_k)=\frac{\log M_q(k,L;v)-q\log M_1(k,L;v)}{1-q}
\eeq
where we introduce the moment sum
\beq
M_q(k,L;v)=\sum_{j=1}^v |\sec(k L_j)|^{2q}.
\eeq

\section{Semiclassical Multifractality}

\subsection{Semiclassical asymptotics of the Renyi entropy}

Let us suppose that the eigenvalue $k$ is in the $m$th cluster and that $s_{m,i+1}<k<s_{m,i}$. 

We have
\beq
\begin{split}
\cos(k L_j)&=(-1)^{m+1}\sin((k-s_{m,j})L_j)\\
&= (-1)^{m+1}(k-s_{m,j})L_j+O(\frac{m^3\epsilon^3\ell_v^3}{L^3}).
\end{split}
\eeq

Therefore, we have, as $v\to\infty$,
$$M_q(n,L;v)\asymp\sum_{j=1}^v(k-s_{m,j})^{-2q}L_j^{-2q}.$$

Moreover, the spacing is of order
$$|s_{m,j}-s_{m,k}|\asymp \frac{m\epsilon}{L^2}|\ell_j-\ell_k|.$$
Hence, we renormalize: $k'=kL^2/(m\epsilon)$, $s'_{m,j}=s_{m,j}L^2/(m\epsilon)$, so that we have $|s'_{m,j}-s'_{m,k}|\asymp |\ell_j-\ell_k|$. And, since $L_j\asymp L$, we have
$$M_q(n,L;v)\asymp (\frac{m\epsilon}{L^2})^{-2q}\sum_{j=1}^v|k'-s'_{m,j}|^{-2q}.$$

%For simplicity take $i=1$. 
We have
$k'-s'_{m,j}=(k'-s'_{m,i})+(s'_{m,i}-s'_{m,j})$
and assume we have a subsequence where $\sigma_m=k'-s'_{m,i}\to\sigma>0$ as $m\to+\infty$ along this subsequence.
So, it is sufficient to study the sum
$$\sum_{j=1}^v|\sigma_m+\ell_i-\ell_j|^{-2q}=\sum_{j\leq i+2}|\sigma_m+\ell_i-\ell_j|^{-2q}+\sum_{j>i+2}(\ell_j-\ell_i-\sigma_m)^{-2q}.$$

We have
$$\sum_{j>i+2}(\ell_j-\ell_i)^{-2q}\leq\sum_{j>i+2}(\ell_j-\ell_i-\sigma_m)^{-2q}\leq \sum_{j>i+2}(\ell_j-\ell_{i+1})^{-2q}.$$

Let us introduce an invertible function $\j\in C^1(\R_+)$ which satisfies $\j(\ell_j)=j$, $\forall j=1,\dots,v$.
We then have
$$\sum_{j=i+3}^v(\ell_j-\ell_{i+1})^{-2q}\leq \int_{i+2}^{v}(\j^{-1}(t)-\ell_{i+1})^{-2q}dt$$
and
$$\sum_{j=i+3}^v(\ell_j-\ell_i)^{-2q}\geq \int_{i+3}^{v+1}(\j^{-1}(t)-\ell_i)^{-2q}dt.$$

Define the Mellin transform of a compactly supported piecewise continuous function $f$ as
$$(\cM f)(s)=\int_0^\infty t^{-s}f(t)dt.$$

We may rewrite 
$$\int_{i+2}^{v}(\j^{-1}(t)-\ell_{i+1})^{-2q}dt=\int_{\ell{i+2}-\ell_{i+1}}^{\ell_v-\ell_{i+1}}t^{-2q}\j'(t+\ell_{i+1})dt
=(\cM F(v,\j))(2q)$$ 
where $$F(v,\j)(t)=(\mathbbm{1}_{[\ell_{i+2},\ell_v]}\j')(t+\ell_{i+1}).$$

If $$\lim_{v\to+\infty}\frac{\log (\cM F(v+1,\j))(s)}{\log (\cM F(v,\j))(s)}=1,$$ then the asymptotics of the Renyi entropy may be calculated in terms of the Mellin transform as 
\begin{equation}
\begin{split}
H_q(\mu_k)=&\frac{\log M_q(k,L;v)-q\log M_1(k,L;v)}{1-q}\\
\sim&\frac{\log (\cM F(v,\j))(2q)-q\log (\cM F(v,\j))(2)}{1-q}.
\end{split}
\end{equation}

\subsection{An example of multifractality}
Let us consider $\ell_j=j^{1/n}$ for some integer $n>0$. Then $\j(t)=t^n$. Let us suppose $q>n/2$. Set $\sigma_1=1+\sigma$. 

Asymptotically, the Mellin transform can be evaluated as
\beq
\begin{split}
(\cM F(v,\j))(2q)
=&\int_{2^{1/n}}^{v^{1/n}}(s-\sigma_1)^{-2q}ns^{n-1}ds\\
=&n\int_{2^{1/n}-\sigma_1}^{v^{1/n}-\sigma_1}s^{-2q}(s+\sigma_1)^{n-1}ds\\
=&n\sum_{k=0}^{n-1}c_{n,k}\sigma_1^{n-1-k}\int_{2^{1/n}-\sigma_1}^{v^{1/n}-\sigma_1}s^{-2q+k}ds\\
%=&n\sum_{k=0}^{n-1}c_{n,k}\sigma_1^{n-1-k}\frac{1}{-2q+k+1}\left[s^{-2q+k+1}\right]_{2^{1/n}-\sigma_1}^{v^{1/n}-\sigma_1}\\
=&n\sum_{k=0}^{n-1}c_{n,k}\sigma_1^{n-1-k}\frac{(2^{1/n}-\sigma_1)^{-2q+k+1}}{-2q+k+1}+O(v^{1-2q/n})
\end{split}
\eeq

For $n\geq 2$, we have
\beq
(\cM F(v,\j))(2)=
\int_{2}^v (t^{1/n}-1-\sigma)^{-2}dt\sim
\begin{cases}
\frac{n}{n-2}v^{1-2/n}, \quad \text{if}\;n\geq 3,\\
\\
\log v, \quad \text{if}\;n=2.
\end{cases}
\eeq

This yields the following asymptotics for the Renyi entropy:
\beq
H_q(k_n)=\frac{\log M_q -q\log M_1}{1-q}\sim
\begin{cases}
\frac{q}{q-1}(1-\frac{2}{n})\log v, \quad \text{if}\;n\geq 3,\\
\\
\frac{q}{q-1}\log\log v, \quad \text{if}\;n=2.\\
\end{cases}
\eeq

We note that the dependence on $\sigma$ only appears in lower order terms.  In particular, for $q>n/2$ and $n\geq 2$, we have the fractal exponent
\beq
D_q=\lim_{k\to+\infty}\frac{H_q(k)}{\log v}=
\begin{cases}
1-\frac{2q-n}{n(q-1)}, \quad\text{if}\; n\geq3,\\
\\
0, \quad\text{if}\; n=2,
\end{cases}
\eeq
where we used $v\to+\infty$, as $k\to+\infty$.

For an example where $D_q=1$, consider $\ell_j=(v+j)^\delta$, $\delta>0$, which leads to moment sums of the form
$$\sum_{j=1}^v(\sigma'+(v+j)^\delta)^{-2q}\asymp v^{1-2\delta q}.$$

So $\log M_q\sim(1-2q)\delta\log v$, which yields the asymptotic
$$H_q(k)=\frac{\log M_q-q\log M_1}{1-q}\sim \frac{(1-2\delta q-q(1-2\delta))\log v}{1-q}=\log v,$$
and, hence, $D_q=\lim_{m\to+\infty}H_q(k)/\log v=1$.

\section{Ground state multifractality}

Let us now fix $m=0$ and take $\epsilon=\epsilon(v)$ to be a decaying function such that $\epsilon\ell_v\to0$, which ensures separate clusters.

Note that this regime is markedly different from the semiclassical regime above. In the semiclassical regime we considered a sequence of eigenvalues which tends to infinity ($m\to+\infty$) and the number of vertices $v=v(m)\to+\infty$ is coupled to the semiclassical parameter to ensure separation of the clusters.

In the ground state regime we consider the lowest eigenvalue in the bottom cluster in a weak coupling regime such that $k_0=\pi/(2L)+\sigma_0$ and $\sigma_0\to0$. In this regime we have $m=0$ and, therefore, $m$ is not coupled to $v$. Analogously to our definition for the ground state of \u{S}eba's billiard we, therefore, define the fractal exponent as 
\beq
D_{q,v}=\lim_{\sigma_0\to0}\frac{\log M_q(k_0)-q \log M_1(k_0)}{1-q}.
\eeq

The moment sum for $\cC_0$ is of the form
$$M_{q}(k_0)=\sum_{j=1}^v(k_0-s_{0,j})^{-2q}=\sum_{j=1}^v\left(\sigma_0+\frac{\pi\epsilon(v)\ell_j}{2L(L+\epsilon\ell_j)}\right)^{-2q}$$
so that
$$\lim_{\sigma_0\to0}M_{q}(k_0)=\left(\frac{\pi\epsilon(v)}{2L}\right)^{-2q}\sum_{j=1}^v\left(\frac{\ell_j}{L+\epsilon\ell_j}\right)^{-2q}$$
and, in turn, we compute
\beq
D_{q,v}=\frac{\log\zeta_{G,v}(2q)-q\log\zeta_{G,v}(2)}{1-q}
\eeq
where 
\beq 
\zeta_{G,v}(s)=\sum_{j=1}^v\left(\frac{\ell_j}{L+\epsilon\ell_j}\right)^{-s}.
\eeq

Now, since $m$ is fixed, we may take $v\to\infty$ which gives rise to the zeta function associated with the length spectrum of the graph
\beq
\zeta_{G,\infty}(2q)=\sum_{j=1}^\infty\ell_j^{-2q}
\eeq

For example take $\ell_j=j$, then the arithmetic graph $G_\N$ which arises in the limit $v\to+\infty$ has length spectrum $\N$ and gives rise to the Riemann zeta function
\beq
\zeta_{G_\N,\infty}(2q)=\zeta(2q)=\sum_{n=1}^\infty n^{-2q}.
\eeq

If one takes the set of lengths $\ell_j$ to be the set of lengths of vectors in a rectangular lattice $\cL$, then the associated graph $G_\cL$ gives rise to the Epstein zeta function
\beq
\zeta_\cL(2q)=\sum_{\xi\in\cL\setminus\{0\}}\|\xi\|^{-2q}=\zeta_Q(q)
\eeq 
where $Q$ is the quadratic form induced by the Euclidean norm $\|\cdot\|$.

If we recall the functional equation for Riemann's zeta function 
$$\zeta(1-s)=\varphi(s)\zeta(s)$$
then we have the following symmetry relation of the fractal exponent $D_q$ around the value $q=1/4$ under the transformation $q\to 1/2-q$:
\beq
\begin{split} 
D_{1/2-q}=&\frac{\log\zeta(2(1/2-q))-(1/2-q)\log\zeta(2)}{1-(1/2-q)}\\
=&\frac{\log \varphi(2q) +\log\zeta(2q)+(q-1/2)\log\zeta(2)}{1/2+q}\\
=&D_q\times\frac{1-q}{1/2+q}+\frac{\log\varphi(2q)+(2q-1/2)\log\zeta(2)}{1/2+q}
\end{split}
\eeq

\end{document}